\documentclass[journal=jcisd8, manuscript=article]{achemso} %add layout=twocolumn
\usepackage{amsmath, amsfonts, amssymb, graphicx, subfigure, xcolor, soul,indentfirst}
\usepackage[labelfont=bf]{caption} % make Fig. be bold

\author{Hao Tian}
\affiliation{Department of Chemistry, Center for Research Computing, Center for Drug Discovery, Design, and Delivery (CD4), Southern Methodist University, Dallas, Texas, United States of America}

\author{Peng Tao}
\affiliation{Department of Chemistry, Center for Research Computing, Center for Drug Discovery, Design, and Delivery (CD4), Southern Methodist University, Dallas, Texas, United States of America}
\email{ptao@smu.edu}

\title{ivis Dimensionality Reduction Framework for Biomacromolecular Simulations}

\begin{document}

\begin{tocentry} %graphical TOC
\includegraphics{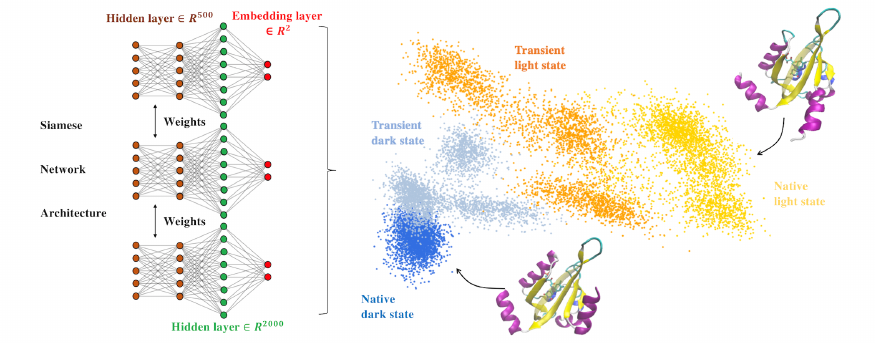}
\\

Title: ivis Dimensionality Reduction Framework for Biomacromolecular Simulations

Authors: Hao Tian and Peng Tao
\end{tocentry}

\begin{abstract}

Molecular dynamics (MD) simulations have been widely applied to study macromolecules including proteins. However, high-dimensionality of the datasets produced by simulations makes it difficult for thorough analysis, and further hinders a deeper understanding of biomacromolecules. To gain more insights into the protein structure-function relations, appropriate dimensionality reduction methods are needed to project simulations onto low-dimensional spaces. Linear dimensionality reduction methods, such as principal component analysis (PCA) and time-structure based independent component analysis (t-ICA), could not preserve sufficient structural information. Though better than linear methods, nonlinear methods, such as t-distributed stochastic neighbor embedding (t-SNE), still suffer from the limitations in avoiding system noise and keeping inter-cluster relations. ivis is a novel deep learning-based dimensionality reduction method originally developed for single-cell datasets. Here we applied this framework for the study of light, oxygen and voltage (LOV) domain of diatom \emph{Phaeodactylum tricornutum} aureochrome 1a (PtAu1a). Compared with other methods, ivis is shown to be superior in constructing Markov state model (MSM), preserving information of both local and global distances and maintaining similarity between high dimension and low dimension with the least information loss. Moreover, ivis framework is capable of providing new prospective for deciphering residue-level protein allostery through the feature weights in the neural network. Overall, ivis is a promising member in the analysis toolbox for proteins.

\end{abstract}

\section{Introduction}

Molecular dynamics (MD) simulations have been widely used in biomolecules to provide insights into their functions at the atomic-scale mechanisms. \cite{lindorff2011fast} For this purpose, extensive timescale is generally preferred for the simulations to study protein dynamics and functions. Due to the arising of graphics processing units (GPU) and their application in biomolecular simulations, MD simulation timescale has reached from nanoseconds to experimentally meaningful microseconds. \cite{shaw2009millisecond,eastman2013openmm} However, simulation data for biomacromolecues such as proteins are high-dimensional and suffer from the curse of dimensionality \cite{indyk1998approximate}, which hinders in-depth analysis, including extracting slow time-scale protein motions \cite{ichiye1991collective}, identifying representative protein structures \cite{zhou2019allosteric} and clustering kinetically similar macrostates \cite{freddolino2008ten}. In order to make these analyses feasible, it will be informative to construct a low-dimensional space to characterize protein dynamics in the best way possible. 

In recent years, new dimensionality reduction algorithms have been developed and can be applied to analyze protein simulations, construct representative distribution in low dimensional space, and extract intrinsic relations between protein structure and functional dynamics. These methods can be generally categorized into linear and nonlinear methods \cite{roweis2000nonlinear,tenenbaum2000global}. Linear dimensionality reduction methods produce new variables as the linear combination of the input variables, such as principal component analysis (PCA) \cite{levy1984quasi} and time-structure based independent component analysis (t-ICA) \cite{naritomi2011slow}. Nonlinear methods construct variables through a nonlinear function, including t-distributed stochastic neighbor embedding (t-SNE) \cite{maaten2008visualizing} and auto encoders \cite{hinton2006reducing}. It is reported that nonlinear methods are more powerful in reducing dimensionality while preserving representative structures \cite{ferguson2011nonlinear}. 

Information is inevitably lost to certain degree through the dimensionality reduction process. \cite{zhao2010multi} It is expected that the distances among data points in the low dimensional space resemble the original data in the high dimensional space. Markov state model (MSM) is often applied to study the dynamics of biomolecular system. MSM is constructed by clustering states in the reduced dimensional space to catch long-time kinetic information. \cite{suarez2016accurate} However, many dimensionality reduction methods, such as PCA and t-ICA, fail to keep the similarity characteristics in the low dimension, which would cause a misleading clustering analysis based on the projections of low-dimensional space. \cite{doerr2017dimensionality} Therefore, more appropriate dimensionality reduction methods are needed to build proper MSM. 

A novel framework, ivis \cite{szubert2019structure}, is a recently developed dimensionality reduction method for single-cell datasets. ivis is a nonlinear method based on siamese neural networks (SNNs) \cite{koch2015siamese}. The SNN architecture consists of three identical neural networks and ranks the similarity to the input data. The loss function used for training process is a triplet loss function \cite{hermans2017defense} that calculates the Euclidean distance among data points and simultaneously minimizes the distances between data of the same labels while maximizing the distances between data of different labels. Due to this intrinsic property, ivis framework is capable of preserving both local and global structures in low-dimensional surface.

\begin{figure*}[t]
 \includegraphics[width=0.98\textwidth]{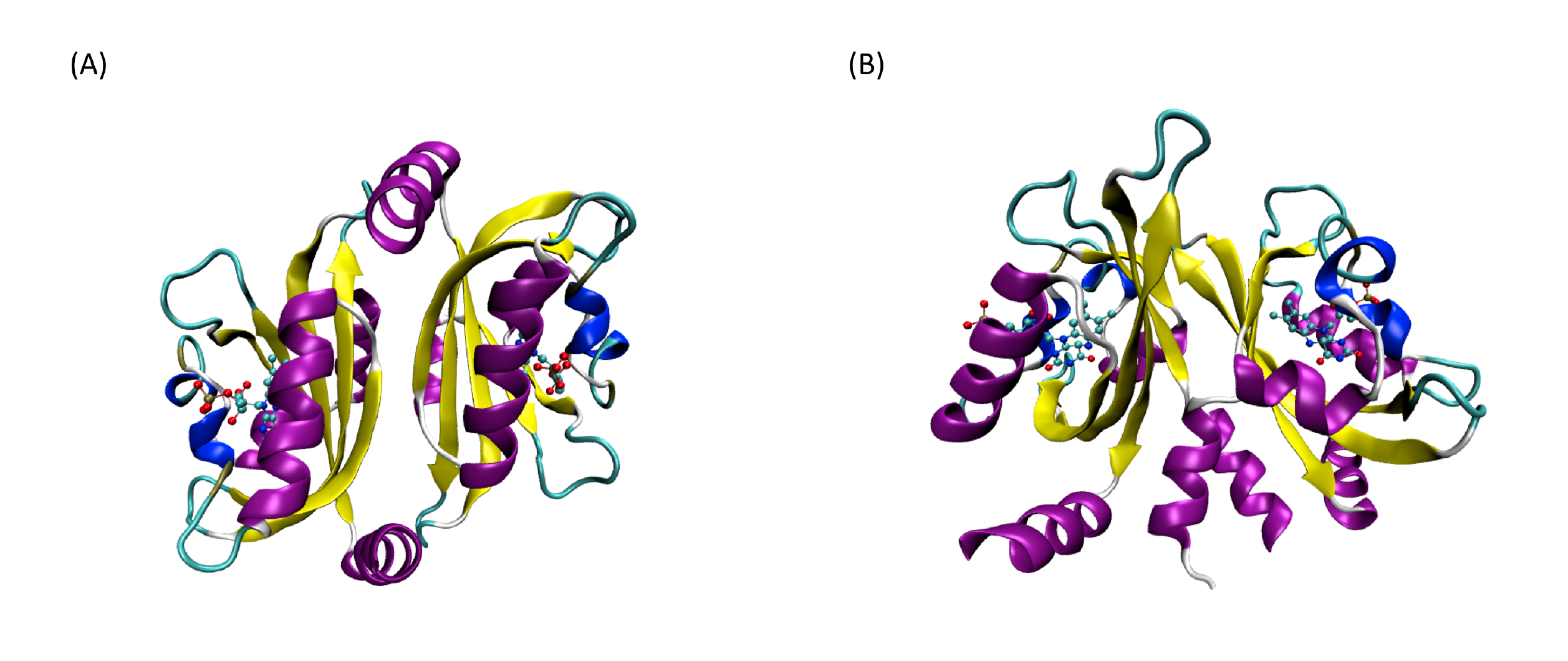}
 \caption{Native structures of AuLOV. (A) Two monomers in the native dark state; (B) A dimer in the native light state. The sequence of secondary structure starts from Ser240 to Glu367.}
 \label{structure}
\end{figure*}

With the success in single-cell expression data, ivis framework is promising as a dimensionality reduction method for simulations of biomacromolecules to investigate their functional dynamics such as allostery. Diatom \emph{Phaeodactylum tricornutum} aureochrome 1a (PtAu1a) is a recently discovered light, oxygen, or voltage (LOV) protein from the photosynthetic stramenopile alga Vaucheria frigida. \cite{takahashi2007aureochrome} This protein consists of an N-terminal domain, a C-terminal LOV core, and a basic region leucine zipper (bZIP) DNA-binding domain. PtAu1a is a monomer in the native dark state. The interaction between its LOV core and bZIP prohibits DNA binding. \cite{heintz2016blue} Upon light perturbation, a covalent bond forms between C4a position of cofator flavin monocleotide (FMN) and sulfur in cysteine 287, triggering a conformational change that leads to the LOV domain dimerization. In the current study, the PtAu1a LOV domain (AuLOV) with two flanking helices (A'$\alpha$ and J$\alpha$ helices) are simulated through MD simulations. The structures of both dark state and light state are shown in Figure \ref{structure}. The main difference between AuLOV and most other LOV proteins is that LOV domain lies in the C-terminal in AuLOV while in the N-terminal in other LOV proteins. \cite{losi2008bacterial, crosson2003lov} Therefore, the conformational changes in AuLOV are expected to differ from other LOV protein, raising the question on how the allosteric signal transmits in AuLOV. In the current study, ivis framework, together with other dimensionality reduction methods, is applied to project the AuLOV simulations onto reduced dimensional spaces. The performance of the selected methods are assessed and compared, validating the ivis as a superior framework for dimensionality reduction of biomacromolecule simulations.

\section{Methods}

\subsection{Molecular Dynamics (MD) simulations}

The crystal structures of AuLOV dark and light states were obtained from Protein DataBank (PDB) \cite{berman2000protein} with PDB ID 5dkk and 5dkl, respectively. The light structure sequence starts from Gly234 while the dark structure sequence starts from Phe239 in chain A and Ser240 in chain B. For consistency, residues before Ser240 were removed to keep the same number of residues in all chains. Therefore, simulations of dark state and light state can be treated similarly. Both structures contain FMN as cofactor. The FMN force field from a previous study \cite{freddolino2013signaling} was used in this study. Two new states, named as transient dark state (forcing the Cysteinyl-Flavin C4a adduct in the dark state structure) and transient light state (breaking the Cysteinyl-Flavin C4a adduct in the light state structure) were constructed to fully explore the protein conformational space. Two monomers (Figure {\ref{structure}A}) and a dimer (Figure {\ref{structure}B}) were simulated in the dark states and light states, respectively.

The crystal structures with added hydrogen atoms were solvated within a rectangular water box using the TIP3P water model \cite{jorgensen1983comparison}. Sodium and chlorine ions were added for charge neutralization. Energy minimization was done for each water box. The system was further subjected to $20$ picoseconds ($ps$) of MD simulations to raise temperature from $0K$ to $300K$ and another $20 ps$ simulations for equilibrium. $10$ nanoseconds ($ns$) of isothermal-isobaric ensemble (NPT) MD simulation under $1$ bar pressure were conducted. Canonical ensemble (NVT) is usually applied in the production runs to investigate the allosteric process {\cite{buchenberg2014long,tsuchiya2019autoencoder}}. $1.1$ microseconds ($\mu s$) of canonical ensemble (NVT) Langevin MD simulation at $300K$ was carried out for each production run. The Langevin dynamics friction coefficient that couples the system to heat bath, was set to 1 $ps^{-1}$,{\cite{cerutti2008vulnerability,ben2019allosteric}} with minimum perturbation to the dynamical properties of the protein system.{\cite{braun2019best}} For all production simulations, the first $100 ns$ simulation is treated as equilibration stage and not included in the analysis. For each structure, three independent MD simulations were carried out and a total of $12$ $\mu s$ simulations were used in analysis. All chemical bonds associated with hydrogen atom were constrained with SHAKE method. $2$ femtoseconds ($fs$) step size was used and simulation trajectories were saved for every $100 ps$. Periodic boundary condition (PBC) was applied in simulations. Electrostatic interactions were calculated with particle mesh Ewald (PME) algorithm \cite{essmann1995smooth} and a cutoff of $1.2$ nanometers ($nm$). Simulations were conducted using graphics processing unit accelerated calculations of OpenMM \cite{eastman2010openmm} with CHARMM \cite{brooks2009charmm} simulation package version c41b1 and CHARMM27 force field \cite{foloppe2000all}.

\subsection{Feature Processing}

In MD simulations, protein structures are represented as atom positions in Cartesian coordinates. However, this representation is neither rotation invariant nor feasible for analysis purpose due to the significant number of atoms with total of $3N$ degrees of freedom. In order to represent the protein structures with rotational invariance and essential structural information, pair-wised backbone C$\alpha$ distances were selected to represent the overall protein configuration. Following our previously proposed feature processing method \cite{tian2020deciphering}, distances were encoded as a rectified linear unit (ReLU) \cite{nair2010rectified}-like activation function and further expanded as a vector. 

\begin{equation}
\text{ReLU}(x) = \max{(0, x)}
\end{equation}

\subsection{Dimentionality Reduction Methods}

\subsubsection{ivis}

ivis is a deep learning-based method for structure-preserving dimensionality reduction. This framework is designed using siamese neural networks, which implement a novel architecture to rank similarity among input data. Three identical networks are included in the SNN. Each network consists of three dense layers and an embedding layer. The size of the embedding layer was set to $2$, aiming to project high-dimensional data into a 2D space. Scaled exponential linear units (SELUs) \cite{klambauer2017self} activation function is used in the dense layers,

\begin{equation}
\text{selu}(x)= \lambda
\begin{cases}
    x & \text{ if } x > 0\\
    \alpha \exp{(x)} - \alpha, & \text{ if } x \le 0
\end{cases}
\end{equation}

The LeCun normal distribution is applied to initialize the weights of these layers. For the embedding layer, linear activation function is used, and weights are initialized with Glorot's uniform distribution. In order to avoid overfitting, dropout layers with a default dropout rate of $0.1$ are used for each dense layer. 

A triplet loss function is used as the loss function for training, 

\begin{equation}
L_{\text{tri}} (\theta) = \left[ \sum_{a, p, n} D_{a, p} - \min{(D_{a, n}, D_{p, n})} + m \right]_+
\end{equation}

\noindent where $a$, $p$, $n$ are anchor points, positive points, negative points, respectively. $D$ and $m$ are Euclidean distance and margin, respectively. Anchor points are points of interest. The triplet loss function aims to minimize the distance between anchor points and positive points while maximizing the distances between anchor points and negative points. The distance between positive points and negative points are also taken into account, as shown in $\min{(D_{a, n}, D_{p, n})}$ in the above equation. 

The $k$-nearest neighbors (KNNs) are used to obtain data for the triplet loss function. $k$ is a tuning parameter and is set to $100$. For each round of calculation, one point in the dataset is selected as an anchor. A positive point is randomly selected among the nearest $k$ neighbors around the anchor, and a negative point is randomly selected outside the neighbors. For each training epoch, the triplet selection is updated to maximize the differences in both local and global distances.  

If the date set could be classified into different groups based on their intrinsic properties, ivis can also be used as a supervised learning method by combining the distance-based triplet loss function with a classification loss. Supervision weight is a tuning parameter to control the relative importance of loss function in labeling classification. 

The neural network is trained using Adam optimizer function with a learning rate of $0.001$. Early stopping is a method to prevent overfitting in training neural network and is applied in this study to terminate the training process if loss function does not decrease after $10$ consecutive epochs.

\subsubsection{Time-structure Independent Components Analysis (t-ICA)}

t-ICA method finds the slowest motion or dynamics in molecular simulations and is commonly used as dimensionality reduction method for macromolecular simulations \cite{naritomi2011slow}. For a given $n$-dimensional data, t-ICA is employed by solving the following equation: 

\begin{equation}
\bar{C}F = CKF
\end{equation}

\noindent where $K$ is eigenvalue matrix and $F$ is the eigenvector matrix. $\bar{C}$ is the time lag correlation matrix defined as 

\begin{equation}
\bar{C} = \langle \langle x(t) - \langle x(t)\rangle)^t (x(t + \tau) - \langle x(t)\rangle)\rangle
\end{equation}

The results calculated by t-ICA are linear combinations of input features that are highly autocorrelated.

\subsubsection{Principal Component Analysis (PCA)}

PCA is a method that finds the projection vectors that maximize the variance by conducting an orthogonal linear transformation \cite{levy1984quasi}. In the new coordinate system, the greatest variance of the data lies on the first coordinate and is called the first principal component. Principal components can be solved through the singular value decomposition (SVD) \cite{golub1971singular}. Given data matrix $X$, the covariance matrix can be calculated as:

\begin{equation}
C = X^T X / (n-1)
\end{equation}

\noindent where $n$ is the number of samples. $C$ is a symmetric matrix and can be diagonalized as:

\begin{equation}
C = VLV^T
\end{equation}

\noindent where $V$ is a matrix of eigenvectors and $L$ is a diagonal matrix with eigenvalues $\lambda_i$ in descending order.

\subsubsection{t-Distributed Stochastic Neighbor Embedding (t-SNE)}

t-SNE is a nonlinear dimentionality reduction method that tries to embed similar objects in high dimensions to points close to each other in a low dimension space \cite{maaten2008visualizing}. t-SNE has been demonstrated as a suitable dimensionality reduction method for protein simulations. \cite{zhou2018t} The calculation process consists of two stages. First, conditional probability is calculated to represent the similarity between two objects as:

\begin{equation}
p_{j|i} = \frac{\exp{(-||x_i - x_j||^2} / 2\sigma_i^2)}{\sum_{k \neq i }\exp{(-||x_i - x_k||^2} / 2\sigma_i^2)}
\end{equation}

\noindent where $\sigma_i$ is the bandwidth of the Gaussian kernels. 

While the conditional probability is not symmetric since $p_{j|i}$ is not equal to $p_{i|j}$, the joint probability is defined as:

\begin{equation}
p_{ij} = \frac{p_{j|i} + p_{i|j}}{2N}
\end{equation}

In order to better represent the similarity among objects in the reduced map, the similarity $q_{ij}$ is defined as:

\begin{equation}
q_{ij} = \frac{(1+||y_i - y_j||^2)^{-1}}{\sum_{k \neq i}(1+||y_i - y_k||^2)^{-1}}
\end{equation}

Combined with the joint probability $p_{ij}$ and similarity $q_{ij}$, Kullback\textendash Leibler (KL) divergence is used to determine the coordinates of $y_i$ as:

\begin{equation}
KL(P||Q) = \sum_{i \neq j} p_{ij} \log \frac{p_{ij}}{q_{ij}}
\end{equation}

The KL divergence measures the differences between high-dimensional data and low-dimensional points, which is minimized through gradient descent method.

A drawback of traditional t-SNE method is the slow training time. In order to speed up the computational time of dimensionality reduction process, Multicore t-SNE \cite{Ulyanov2016} is used and abbreviated as t-SNE in this study.

\subsection{Performance Assessment Criteria}

Several assessment criteria are applied to quantify and compare the performance of each dimensionality reduction method.

\subsubsection{Root-Mean-Square Deviation (RMSD)}
The RMSD is used to measure the conformational change in each frame with regard to a reference structure. Given a molecular structure, the RMSD is calculated as:

\begin{equation}
\text{RMSD} = \sqrt{\frac{\sum_{i=1}^N (r_i^0 - Ur_i)^2}{N}}
\end{equation}

\noindent where $r$ is a vector represented in Cartesian coordinates and $r_i^0$ is the $i^{th}$ atom in the reference structure.

\subsubsection{Pearson Correlation Coefficient (PCC)}
Pearson correlation coefficient \cite{benesty2009pearson} reflects the linear correlation between two variables. PCC has been rigorously applied to estimate the linear relation between distances in the original space and the reduced space \cite{adler2010quantifying}. L2 distance, which is also called Euclidean distance, is used for the distance calculation and is shown as follows:

\begin{equation}
d_2(p, q) = \sqrt{\sum_{i=1}^n (p_i - q_i)^2}
\end{equation}

Based on the L2 distance expression, PCC is calculated as:

\begin{equation}
r_{xy} = \frac{\sum_{i=1}^n (x_i - \bar{x})(y_i - \bar{y})}{\sqrt{\sum_{i=1}^n (x_i - \bar{x})} \sqrt{\sum_{i=1}^n (y_i - \bar{y})}}
\end{equation}

\noindent where $n$ is the sample size, $x_i, y_i, \bar{x}, \bar{y}$ are the distances and the mean value of distances, respectively.

\subsubsection{Spearman's Rank-Order Correlation Coefficient}

Spearman's rank-order correlation coefficient is used to quantitatively analyze how well distances between all pairs of points in the original spaces have been preserved in the reduced dimensions. Specifically, Spearman correlation coefficient measures the difference in distance ranking, which is calculated as the following:

\begin{equation}
\rho = 1 - \frac{6\sum d_i^2}{n(n^2 - 1)}
\end{equation}

\noindent where $d_i$ is the difference in paired ranks and $n$ equals the total number of samples.

\subsubsection{Mantel Test}
The Mantel test is a non-parametric method that is originally used in genetics, \cite{diniz2013mantel} which tests the correlation between two distance matrices. A common problem in evaluating the correlation coefficient is that distances are dependent to each other and therefore cannot be determined directly. The Mantel test overcomes this obstacle through permutations of the rows and columns of one of the matrices. The correlation between two matrices is calculated at each permutation. MantelTest GitHub repository \cite{mantel} was used to implement the algorithm.

\subsubsection{Shannon Information Content (IC)}
While chemical information in the original space could be lost to a certain degree in the reduced space, dimensionality reduction methods are expected to keep the maximum information. Shannon information content is applied to test the information preservation in the reduced space, which is defined as:

\begin{equation}
I(x) = -\log_2(P)
\end{equation}

\noindent where $P$ is the probability of a specific event $x$. 

To avoid the possible dependency among different features in the reduced dimensions, original space was reduced to $1$ dimension (1D) to calculate the IC. The values in the 1D was sorted and put into $100$ bins of the same length. The bins were treated as events and the corresponding probabilities were calculated as the ratio of the number of samples in each bin to the total number of samples.

\subsubsection{Markov State Model (MSM)}

Markov state model has been widely used to partition the protein conformational space into kinetically separated macrostates \cite{mcgibbon2014statistical} and estimate relaxation time to construct long-timescale dynamics behavior \cite{zhou2019allosteric}. MSMBuilder {\cite{harrigan2017msmbuilder}} (version 3.8.0) was employed to implement the Markov state model. $k$-Means clustering method was used to cluster $1,000$ microstates. A series of lag time at equal interval was set to calculate the transition matrix. The corresponding second eigenvalue was used to estimate the relaxation timescale, which was calculated as: 

\begin{equation}
t(\tau) = -\frac{\tau}{\ln \lambda_1}
\end{equation}

\noindent where $\lambda_1$ is the second eigenvalue and $\tau$ is the lag time.  

The generalized matrix Rayleigh quotient (GMRQ) {\cite{mcgibbon2015variational}}, generated using the combination of cross-validation and variational approach, was used to assess the effectiveness of MSM on dimensions and dimensionality reduction methods. State decompositions are different through various dimensionality reduction methods. A good method is expected to lead to a Markov state model with larger GMRQ value.

\subsubsection{Machine Learning Methods}

\subsubsection*{Random Forest (RF)}
Random forest \cite{liaw2002classification} is a supervised machine learning method that was used in this study for trajectory states classification. A random forest model consists of multiple decision trees, which are a class of partition algorithm that recursively groups data samples of the same label. Features at each split are selected based on the information gain. A final prediction result of random forest is made from results in each decision tree through voting algorithm. For random forest models at each depth, the number of decision tree was set to $50$. Scikit-learn (version 0.20.1) \cite{pedregosa2011scikit} was used for RF implementation. 

\subsubsection*{Artificial Neural Network (ANN)}
An artificial neural network was used to learn the nonlinear relationships of coordinates on the reduced 2D dimension. An ANN is generally formed with input layer, hidden layer and output layer. In each layer, different neurons (nodes) are assigned and connected with adjacent layer(s). During the training process, input data are fed through the input and hidden layers and prediction results are made in the output layer. For each training step, the error between the predicted result and the actual result is propagated from the output layer back to the input layer, which is also called back propagation \cite{hecht1992theory}, and the weight in every neuron is updated. When there is more than one hidden layer, ANN is also referred as deep neural network (DNN), which requires more computation power. To minimize training cost, only two hidden layers, each with $64$ nodes, were used. Adam optimizer \cite{kingma2014adam} was used for weight optimization. ANN was implemented with Keras (version 2.2.4-tf) \cite{chollet2015keras}.

\section{Results}

\subsection*{Dataset of C$\alpha$ distances represents the protein structures}

\begin{figure*}[t]
 \includegraphics[width=0.9\textwidth]{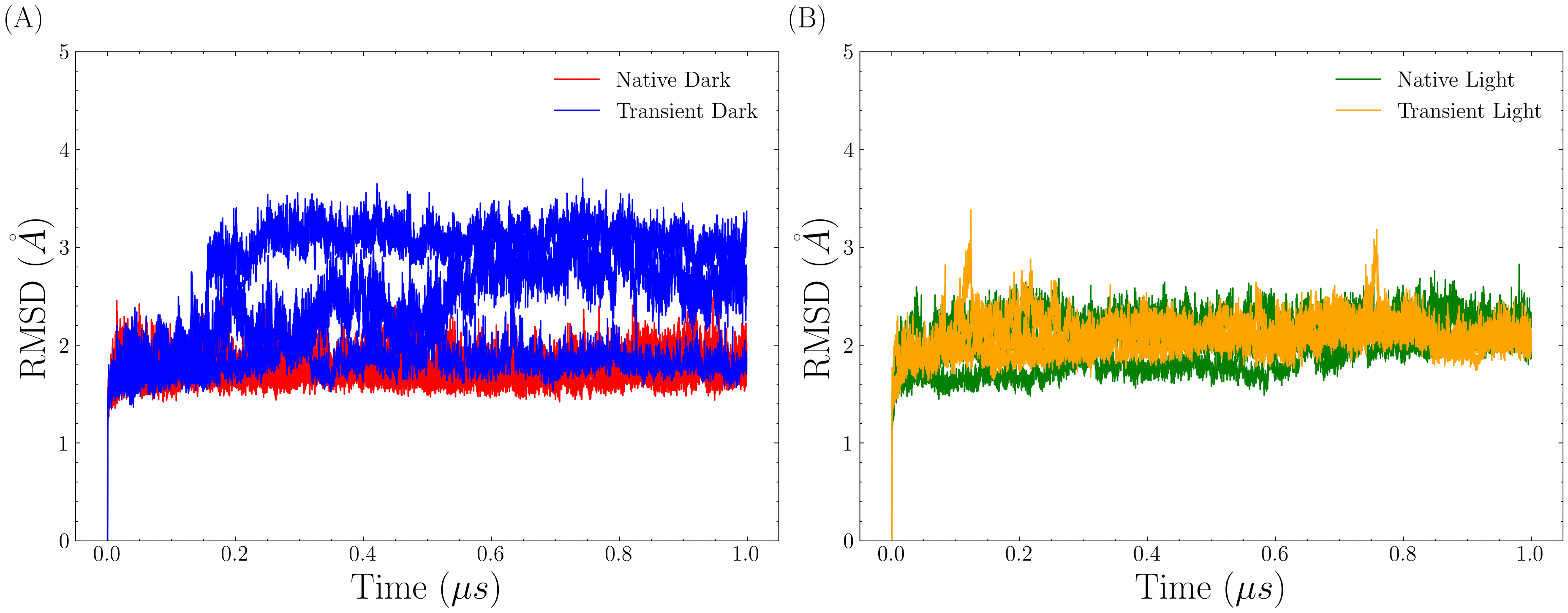}
 \caption{The RMSDs of AuLOV MD trajectories. (A) Native dark and transient dark states; (B) Native light and transient light states. For each state, three independent simulations were carried out. }
 \label{RMSD}
\end{figure*}

There are two native states of AuLOV: native dark state and native light state. To explore the protein response with regard to the formation of the covalent bond between cysteine 287 and FMN, two new states were constructed as transient dark state and transient light state by forcing the Cysteinyl-Flavin C4a adduct in the native dark state and breaking this adduct in the native light state, respectively. The RMSDs of MD simulations are plotted in Figure {\ref{RMSD}}. For each trajectory, the RMSD values were calculated with regard to the first frame. Averaged RMSDs were $1.75\mathring{A}$, $2.04\mathring{A}$, $2.39\mathring{A}$ and $2.08\mathring{A}$ in native dark, transient dark, transient light and native light states, respectively. Compared with the result in native dark state, the higher RMSD value in transient dark state indicates that the light-induced covalent bond Cys287-FMN increases the protein flexibility and dynamics. The transient light state displays the highest averaged RMSD value, indicating the highest flexibility or largest conformational change.

The pair-wised distances of backbone C$\alpha$ in simulations were extracted as features representing the character of protein configurations. There are $254$ residues in the AuLOV structure and total of $ 254 * 253 / 2 = 32,131$ C$\alpha$ distances were calculated. Before further analysis, features were transformed into vectors with our proposed technique outlined in the Methods section. Considering the non-bonded chemical interaction, $10.0 \mathring{A}$ was selected as threshold for feature transformation. There are $10,000$ frames in each trajectory, leading to a sample size of $120,000$ in the overall dataset. Full datasets were applied in all analysis. To gain more statistical significance, each MD trajectory was split into $5$ sub-trajectories at equal intervals. The performance assessments were conducted for each sub-trajectory independently. The mean and standard deviation values of the $5$ subsets were calculated.

\subsection*{Information is well-preserved in ivis dimensionality reduction method}

\begin{figure*}[t]
 \includegraphics[width=0.8\textwidth]{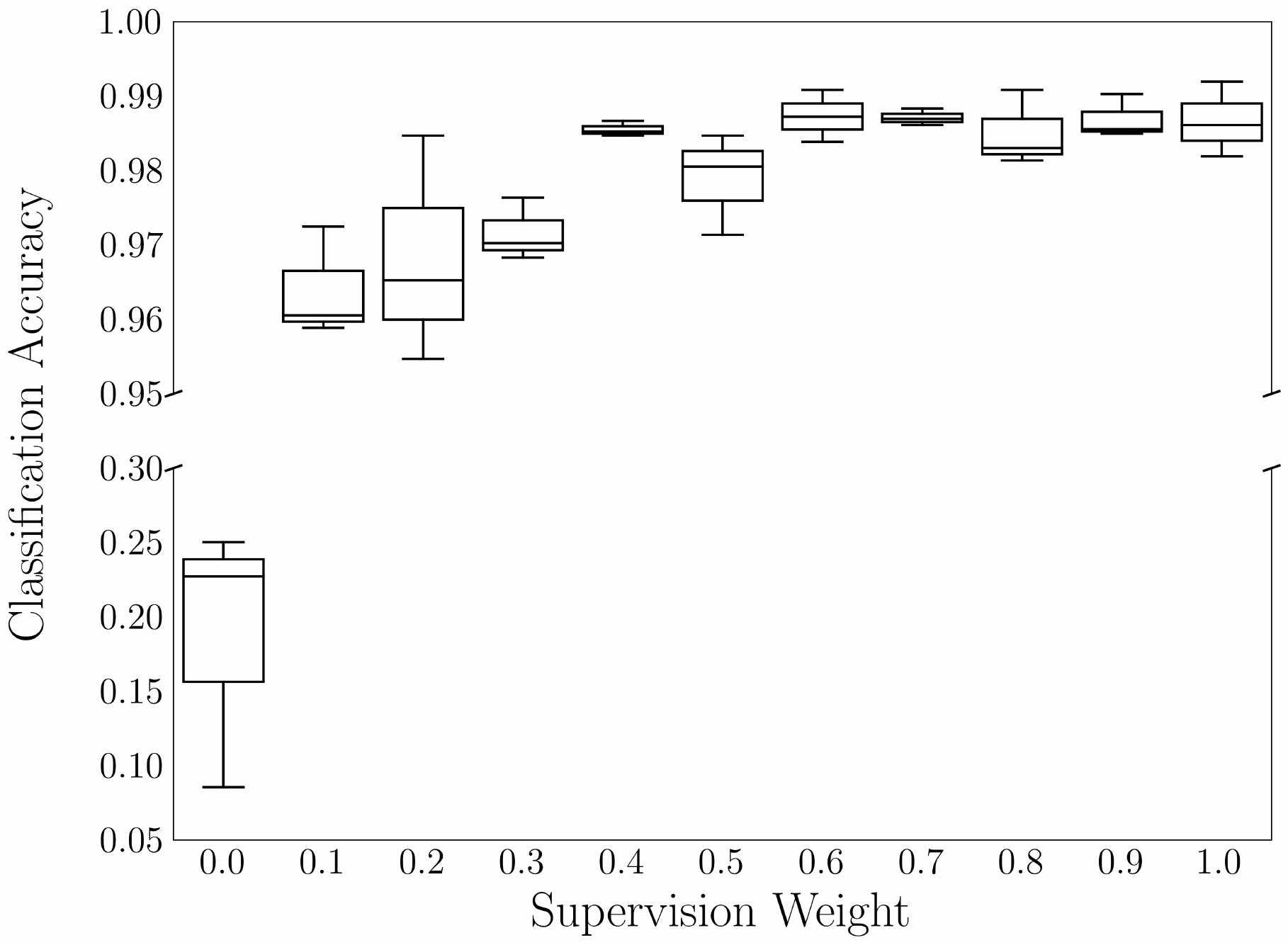}
 \caption{Classification accuracy using ivis framework with different supervision weights. With $0.0$ supervision weight, it is referred to as unsupervised ivis model. Classification accuracy is high for any non-zero supervision weight. Therefore, $0.1$ was chosen as the hyperparameter for supervised ivis. }
 \label{ivis_select_best}
\end{figure*}

Several hyperparameters of ivis model were selected based on the recommended values for different observation sizes. Given the large number of sample size, $k$ was set to $100$ and the number of early stopping epoch was $10$. Maaten neural network architecture was selected, which consists of three dense layers with $500$, $500$ and $2,000$ neurons, respectively. In order to select the best parameter of supervision weight, full trajectory dataset was randomly split into training set ($70\%$) and testing set ($30\%$). ivis models were trained on the training set and validated on the testing set. The prediction result with different supervision weights is plotted in Figure \ref{ivis_select_best}. The ivis model performed poorly at $0.0$ supervision weight, which corresponds to unsupervised ivis, with an average accuracy below $25\%$. The average accuracy values for other supervision weights were stable and over $95\%$. Specifically, there was no significant increase in the accuracy value after $0.1$ supervision weight, which was chosen as the hyperparameter for the supervised ivis model. Unsupervised ivis framework with the same value of other hyperparameters was applied for comparison.

\begin{figure*}[!ht]
 \includegraphics[width=0.99\textwidth]{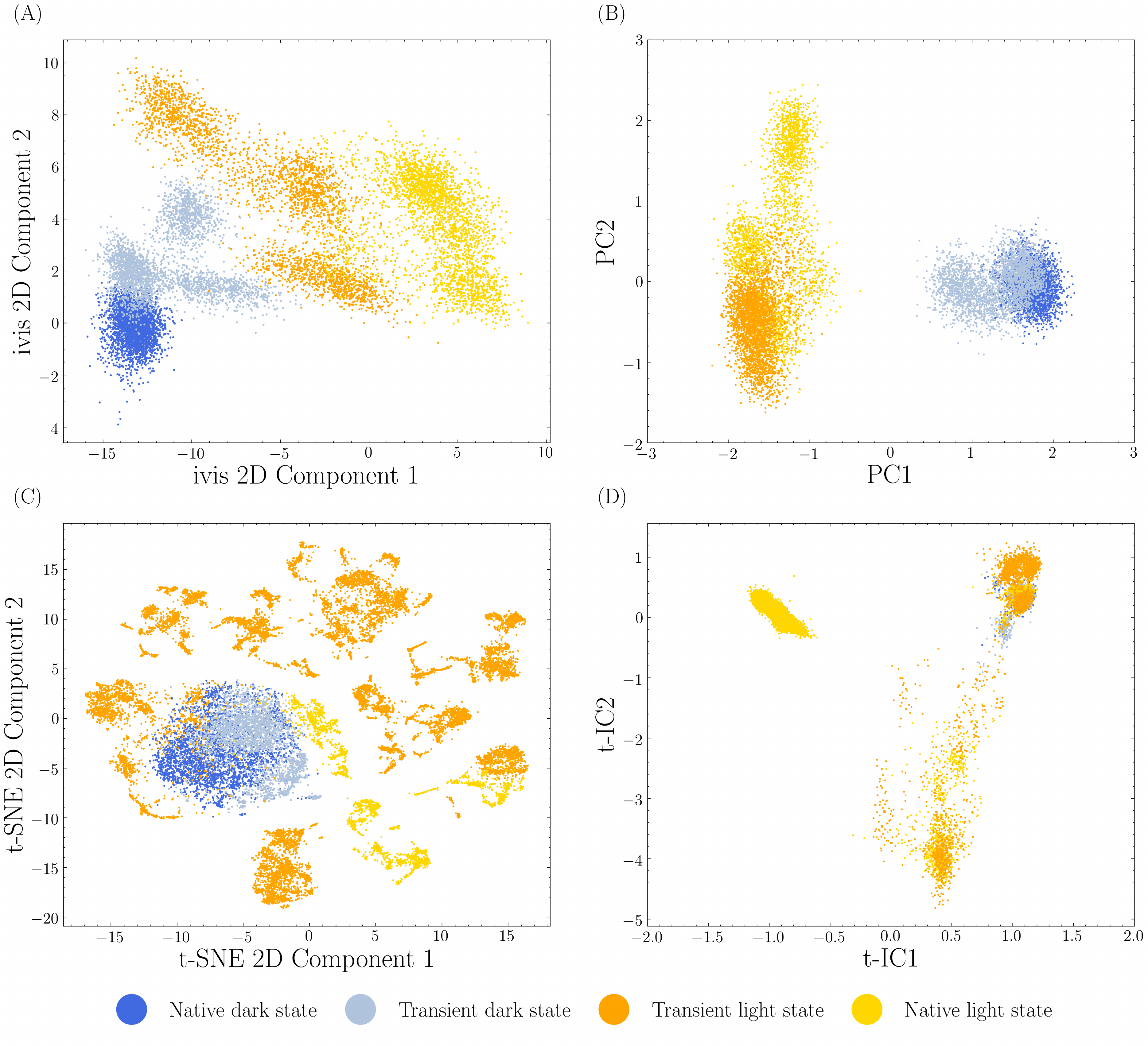}
 \caption{2D projections of four dimensionality reduction methods. (A) Supervised ivis; (B) PCA; (C) t-SNE; (D) t-ICA. }
 \label{2d}
\end{figure*}

Four dimensionality reduction models (supervised ivis, PCA, t-SNE and t-ICA) were applied on the MD simulations to project high dimensional ($32,131$) space to 2D surface (Figure {\ref{2d}}). The supervised ivis dimensionality reduction method, as well as PCA, successfully separated dark and light states while keeping the corresponding transition states close (Figure {\ref{2d}A} and {\ref{2d}B}). These states are important for dynamical analysis as they could be used to reveal the free energy and kinetic transition landscape for target system. For t-SNE (Figure {\ref{2d}C}) and t-ICA (Figure {\ref{2d}D}) projections, transient dark state and native dark state overlap significantly, thus hindering the extraction of thermodynamics and kinetics information. Therefore, supervised ivis dimensionality reduction method and PCA are demonstrated to be proper in representing the chemical information in the low dimension among the investigated methods.

\begin{figure*}[t]
 \includegraphics[width=0.98\textwidth]{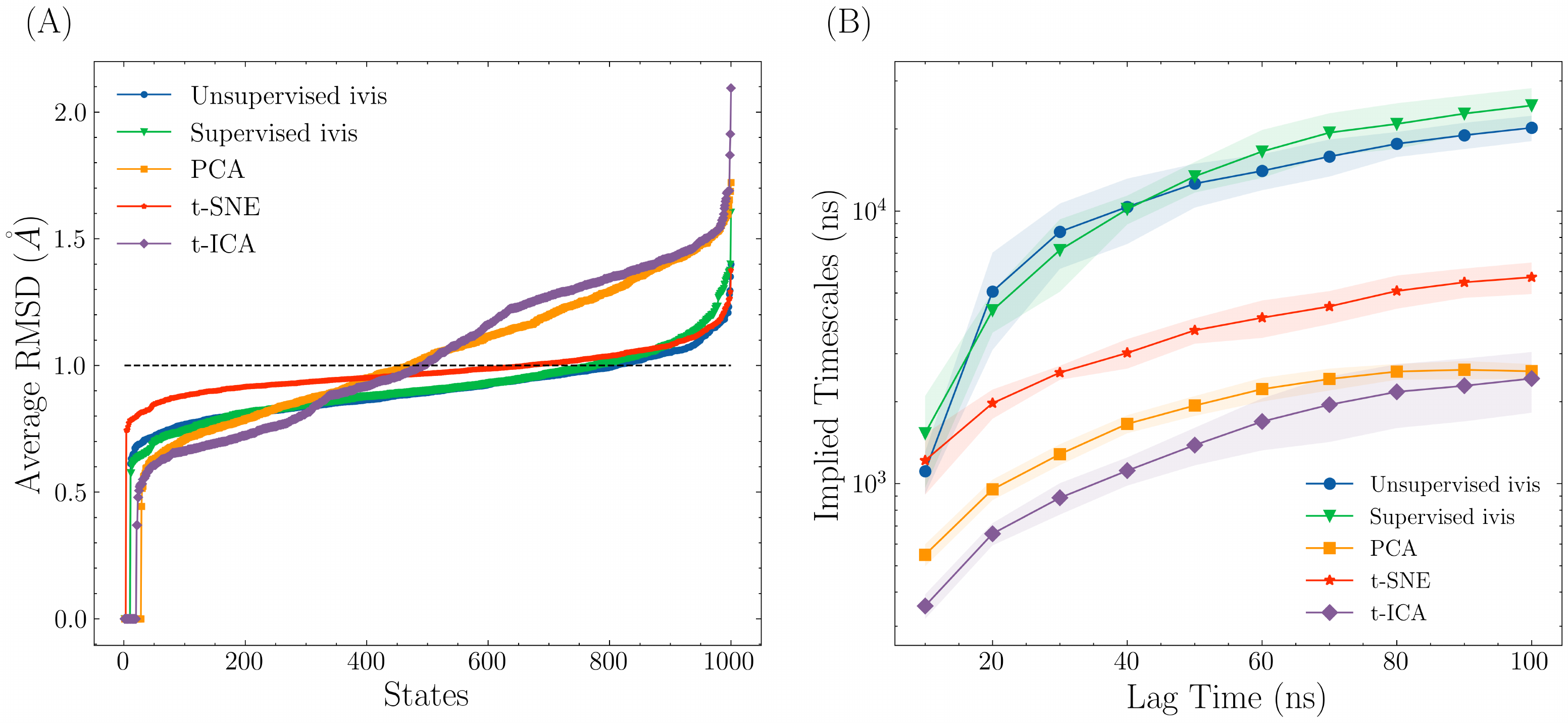}
 \caption{Analysis results of 2D projections for different dimensionality reduction methods. (A) The average values of RMSDs in microstates clustered within projected 2D dimensional space. (B) Estimated implied timescales from Markov state models with regard to different lag times. For each model, the mean value of implied timescale is calculated among five subsets and is plotted in solid color. Standard deviation is calculated to show the stability for each lag time and is illustrated using light color. }
 \label{timescale}
\end{figure*}

The k-means clustering was used in the reduced dimensions to partition a total number of $120,000$ frames from AuLOV MD trajectories into $1,000$ microstates. Within each cluster, the RMSDs were calculated for each structure pair. A RMSD value of each cluster is defined as the average RMSD value among all structure pairs within that cluster. The results of five dimensionality reduction models are shown in Figure \ref{timescale}A. The average RMSD value of an appropriate microstate should be lower than $1.0 \mathring{A}$. \cite{pande2010everything,bowman2009progress} From this prospective, unsupervised ivis and supervised ivis show similar values in each microstate and are the best two methods among the selected methods. As reported previously \cite{zhou2018t}, t-SNE also exhibited good performance in measuring the similarity with the Cartesian coordinates. 

A metric to compare different dimensionality reduction methods is the implied relaxation timescale calculated from Markov state model. To build MSM, MD simulations were projected onto 2D space and $1,000$ microstates were sampled through k-means with corresponding estimated relaxation timescales. For each method, the slowest timescale in each lag time was extracted based on different lag times ranging from $10$ to $100$ ns and is shown in Figure \ref{timescale}B. The convergence of timescales is important for eigenvalues and eigenvectors calculation. \cite{chodera2007automatic} For all five models, relaxation timescales converged, indicating the Markovianity of the MSMs. Both supervised ivis and unsupervised ivis models show long timescales, indicating the effectiveness of MSM built on the reduced spaces.

\begin{figure*}[tbp]
\centering
 \includegraphics[width=0.99\textwidth]{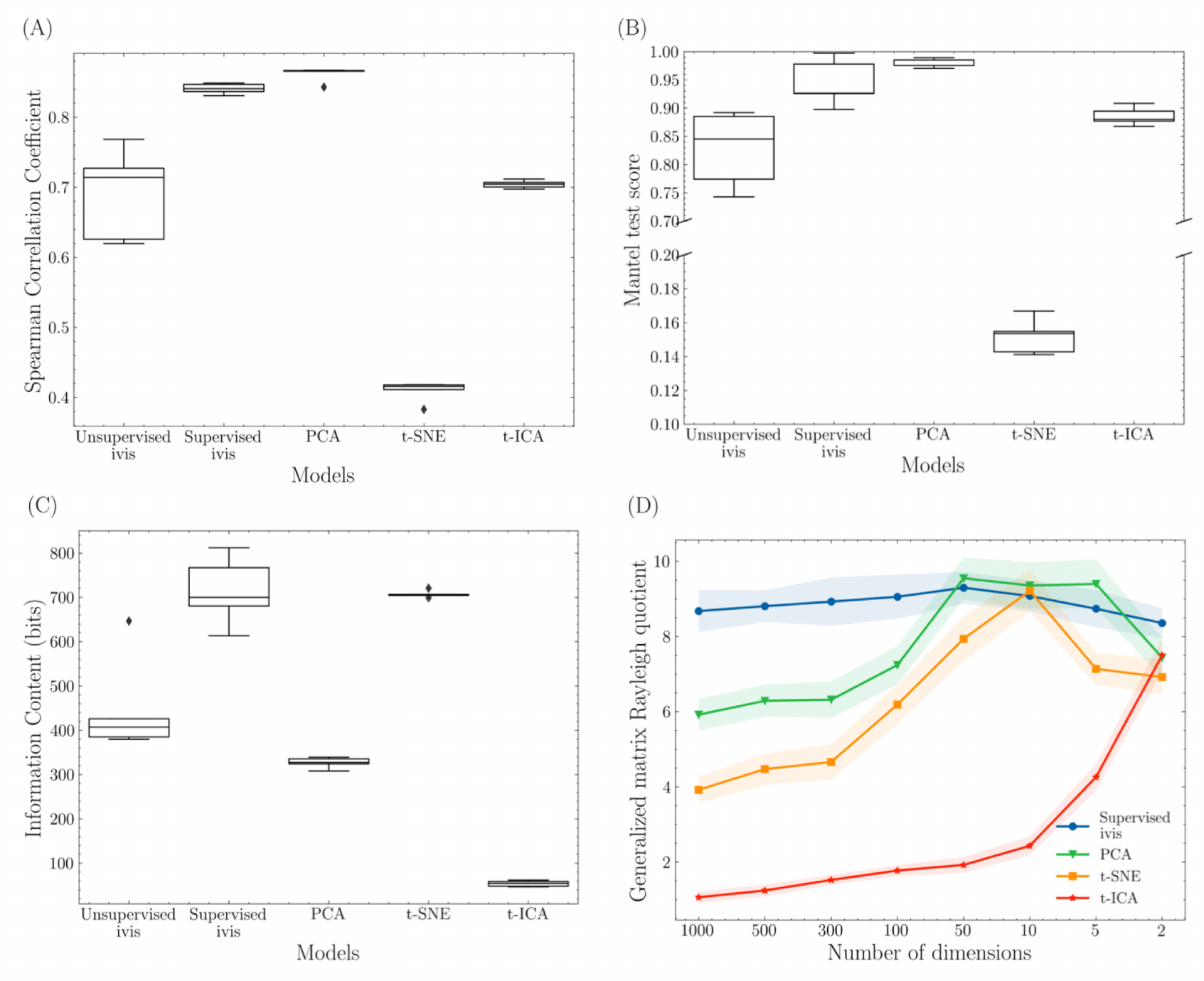}
 \caption{Results of quantitative analysis. (A) Spearman correlation coefficient results of different dimensionality reduction methods. The mean values for unsupervised ivis, supervised ivis, PCA, t-SNE and t-ICA are $0.69$, $0.84$, $0.86$, $0.41$ and $0.70$, respectively. The height of each box represents the interquartile range. (B) The Mantel test scores of different dimensionality reduction methods. The mean values for unsupervised ivis, supervised ivis, PCA, t-SNE and t-ICA are $0.83$, $0.95$, $0.98$, $0.15$ and $0.89$, respectively. (C) Shannon information content of different dimensionality reduction methods. The mean values for unsupervised ivis, supervised ivis, PCA, t-SNE and t-ICA are $449.0$, $714.6$, $327.0$, $707.3$ and $54.0$, respectively. To avoid dependent variables in information content calculation, high-dimensional C$\alpha$ distances were projected to 1D. (D) Generalized matrix Rayleigh quotient with different dimensions and dimensionality reduction methods. }
 \label{quantana}
\end{figure*}

Euclidean distances between data points in the low dimensional space are expected to reflect the similarity in the high dimension. In order to quantify the degree of this relationship kept in reduced dimensional space, Spearman correlation coefficients were calculated between Euclidean distance pairs in the original space and those in the reduced space. The results are shown in Figure \ref{quantana}A. While PCA preserved the Euclidean distances well with an average of $0.86$ coefficient, supervised ivis model showed a comparable high coefficient of $0.84$. The unsupervised ivis model also exhibited the ability to preserve the linear relationship. The poor performance of t-SNE model may be due to the reason that t-SNE is a nonlinear method and therefore suffers the problem that distance in the high dimensional space is not linearly projected to low dimensional space, as reported in other studies \cite{schubert2017intrinsic,zhou2018using}.

While ivis models showed good ability in keeping the linear projection relation, the Spearman correlation coefficient fails to overcome the problem that features are not independent. The pair-wised distances are subjected to the molecular motion of C$\alpha$ that changing the coordinate of one C$\alpha$ atom would affect the distances related to this atom. Therefore, to address this issue, the Mantel test was used to randomize the Euclidean distances. Permutations of rows and columns in the Euclidean distance matrix were done for $10,000$ times while Pearson correlation coefficient being calculated at each time. The results of the Mantel test are plotted in Figure \ref{quantana}B. Both unsupervised ivis and supervised ivis showed remarkable results in preserving the correspondence relationship in randomized order, at the mean coefficient of $0.83$ and $0.95$, respectively.

During the process of dimensionality reduction, information is inevitably lost to some degree. In order to measure the retaining information through the dimensionality reduction process, the Shannon information is applied to the coordinates in the low dimensional space. However, when dealing with multiple variables, especially for the dependent C$\alpha$ distances, the total Shannon information is not equal to the sum of the Shannon information of each variable. To reduce the computation complexity, high dimensional features were reduced to 1D for calculation and results are plotted in Figure \ref{quantana}C. It shows that supervised ivis model is superior in preserving information content with the least information loss. It is also worth noting that t-SNE showed better performance than the unsupervised ivis model.

To study the performance of Markov state model on dimensions and dimensionality reduction methods, the generalized matrix Rayleigh quotient was calculated for each dimension and method (Figure {\ref{quantana}D}). The results of four methods showed different trends. Supervised ivis and t-ICA methods were the least and most affected by the number of dimensions, respectively. For PCA and t-SNE, the optimal parameter of the number of dimensions is in the tens. 2D is typically used for MSM construction and visualization purpose {\cite{wang2019dynamical,wang2019machine,romano2017gradient}}. With this regard, supervised ivis exhibited the highest GMRQ value.

\subsection*{ivis helps in understanding biological system and allosteric mechanism}

\begin{figure*}[t]
\centering
 \includegraphics[width=0.99\textwidth]{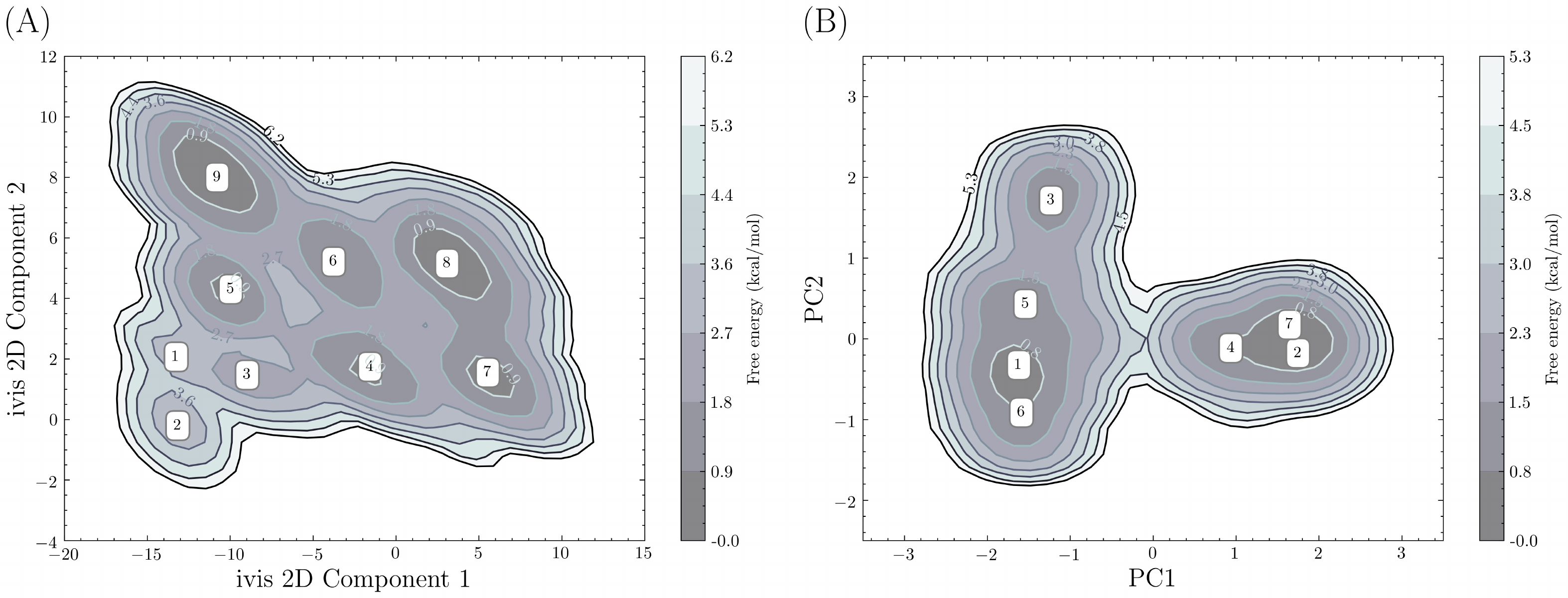}
 \caption{Free energy landscape representations of (A) supervised ivis and (B) PCA projections.}
 \label{free-energy}
\end{figure*}

\begin{figure*}[t]
\centering
 \includegraphics[width=0.9\textwidth]{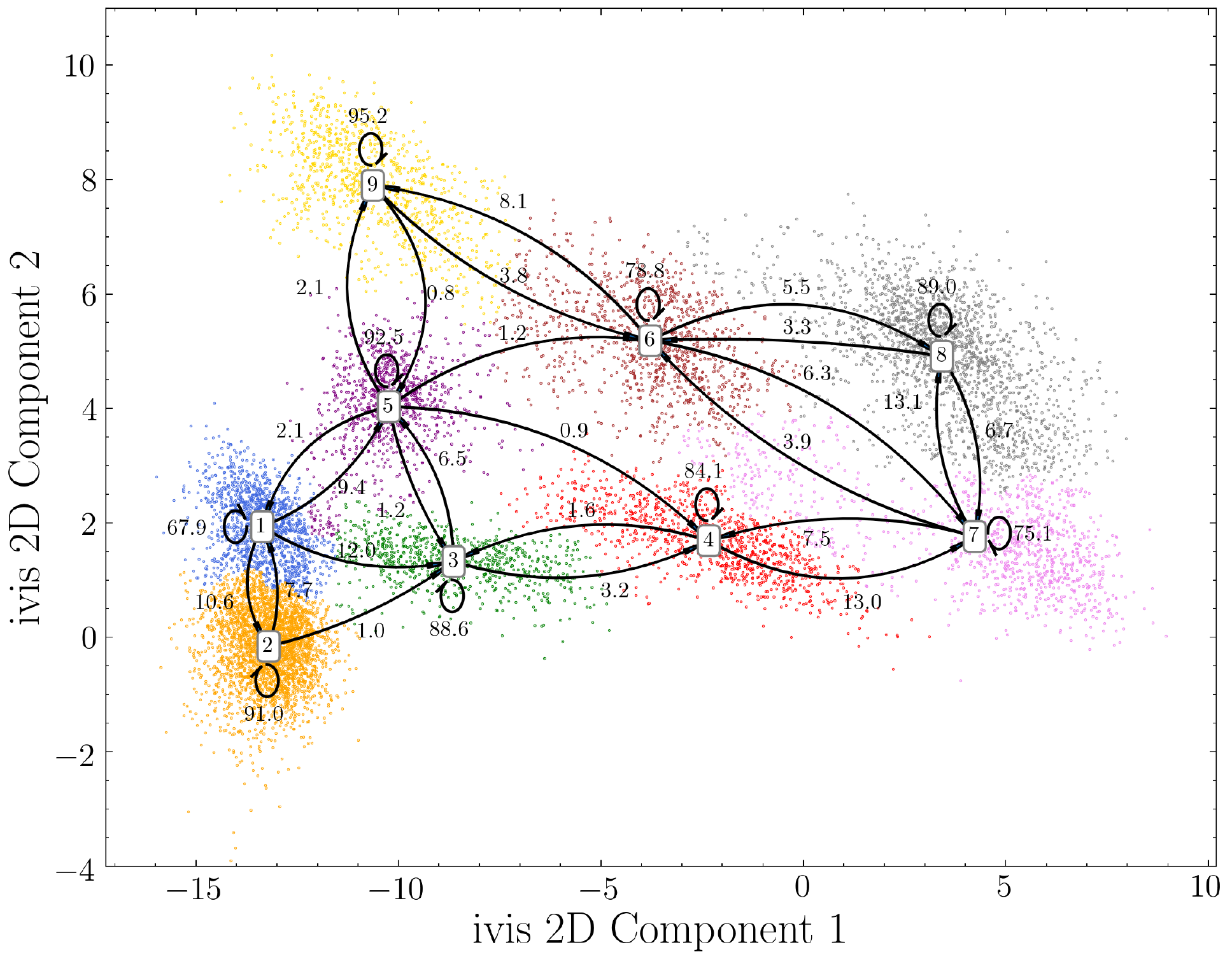}
 \caption{Transition probabilities between macrostates in the ivis dimensionality reduction 2D projections. }
 \label{transition}
\end{figure*}

Based on the results of 2D projections, it is shown that ivis and PCA methods can successfully separate different protein states and reveal the thermodynamics information. To compare the results of these two methods, free energy surfaces were constructed using variables in the 2D projections shown in Figure {\ref{free-energy}}. For each method, the number of macrostates was determined based on the relaxation timescale. There are $9$ and $7$ macrostates in supervised ivis method and PCA, respectively. In supervised ivis result (Figure {\ref{free-energy}}A), each state is represented by separate minimum energy basins while in PCA result (Figure {\ref{free-energy}}B), state $2$ and $7$ share one minimum energy basin, as well as state $1$ and $6$.

The transition probabilities among macrostates in ivis projections are shown in Figure {\ref{transition}}. Based on the similarity with crystal structures, macrostate $2$ and $8$ are referred to as native dark state and native light state, respectively. Other macrostates are considered as transient states. The probabilities of native dark state and native light state to remain to themselves are among the highest ones and indicate high stability of these two states. It is interesting to observe that marcostate $9$ may have the highest stability among all Markov states. Two transient states (macrostate $1$ and $3$) are adjacent to the native dark state and considered vital for allosteric signal propagation upon Cys287-FMN bond formation. 

\begin{figure*}[t]
\centering
 \includegraphics[width=0.99\textwidth]{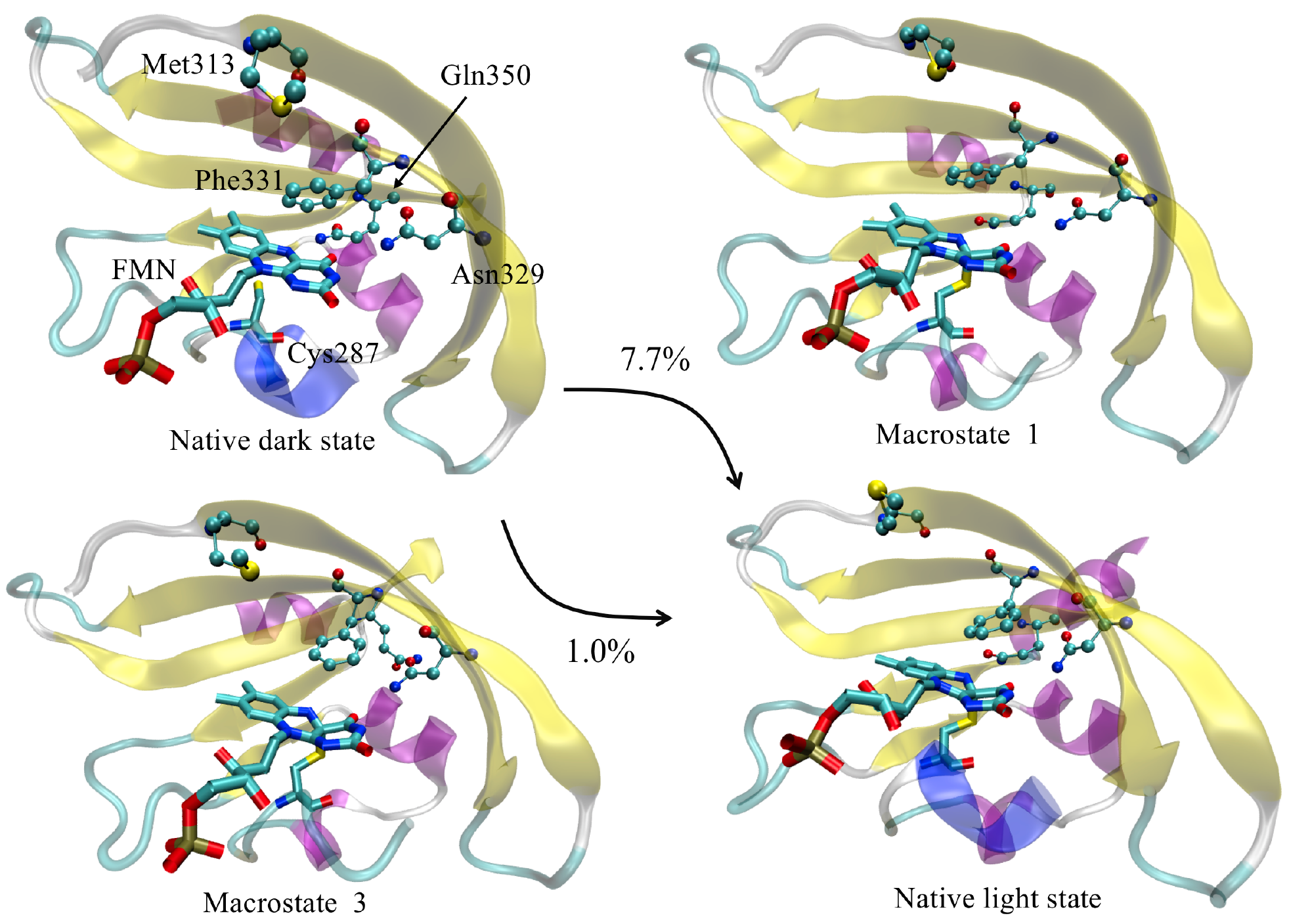}
 \caption{Representative structures of experimentally identified residues near FMN in macrostate $1$ and $3$. Structures of native dark state and native light state are presented for reference. Based on the transition matrix estimated from Markov state model, the transition probabilities of native dark state to macrostate $1$ and mcarostate $3$ are $7.7\%$ and $1.0\%$, respectively. }
 \label{more}
\end{figure*}

Several residues near FMN have been reported to undergo significant conformational changes in the allosteric process {\cite{heintz2016blue}}. Representative structures of these residues in macrostates $1$ and $3$ were extracted and shown in Figure {\ref{more}}. Through the comparison of representative structures in the native dark state and native light state, it is shown that several residues undergo different rotamers' change in macrostates $1$ and $3$. In macrostate $1$, Gln350 and Asn329 display conformational changes by breaking hydrogen bonds with FMN, which is consistent with the orientations in the native light state. In macrostate $3$, Gln350 rotates further from FMN and Phe331 moves closer. The transition probability from native dark state to macrostate $1$ is $7.7\%$ while the transition probability to macrostate $3$ is $1.0\%$. Therefore, starting from the native dark state, the AuLOV allosteric process is more probable to go through macrostate $1$ than macrostate $3$. However, macrostate $1$ is buried in the native dark state of PCA projections, and both macrostates $1$ and $3$ are buried in the native dark state of t-SNE and t-ICA projections, leading to the ambiguity of such comparison using these methods. Thus, ivis framework is proved to be superior in revealing the residue-level mechanism study.

\begin{figure*}[t]
\centering
 \includegraphics[width=0.98\textwidth]{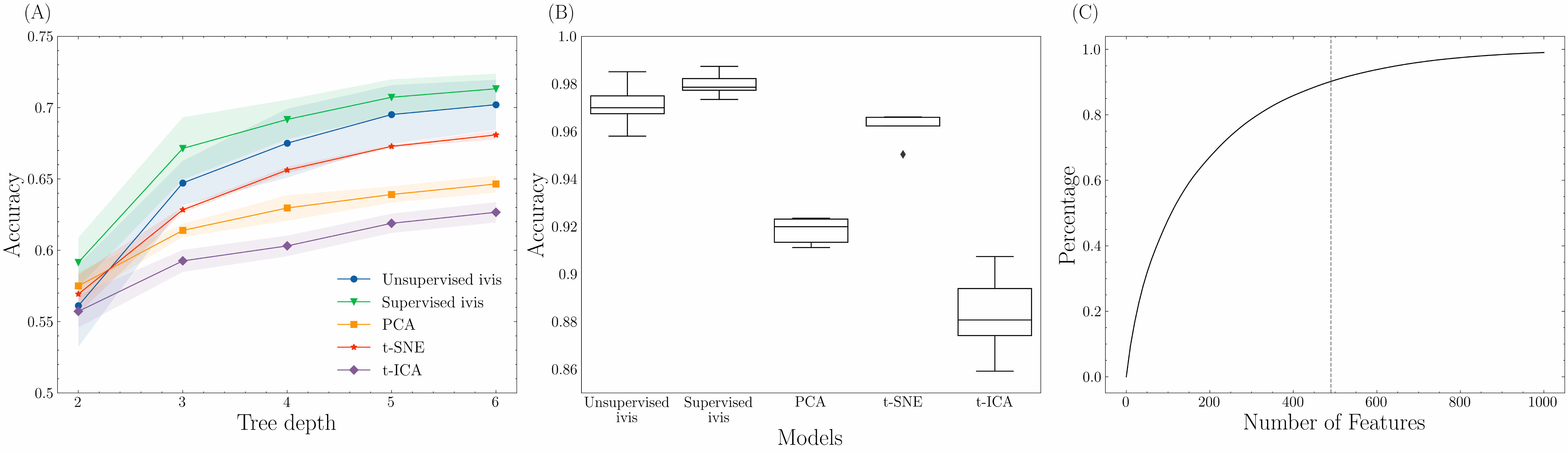}
 \caption{Prediction accuracy of different machine learning models. (A) Random forest and (B) artificial neural network were used on the reduced 2D spaces to predict labels of macrostates from MSM. (C) Accumulated feature importance of random forest model applied in the projections of supervised ivis framework at depth $5$. }
 \label{accuracy}
\end{figure*}

The effectiveness of MSM depends on the projected 2D space, where appropriate discrete states are produced by clustering the original data points in the projection space. The number of macrostates are determined based on the implicated timescales using different lag time in different reduced spaces. In this study, $9$, $9$, $7$, $9$, $7$ macrostates were selected for unsupervised ivis, supervised ivis, PCA, t-SNE and t-ICA, respectively. The samples were clustered through Perron-cluster cluster analysis (PCCA). Dataset was further split into training set ($70\%$) and testing set ($30\%$). Two machine learning methods (random forest and artificial neural network) were applied to predict the macrostates of each data point based on the pair-wised C$\alpha$ distances. Prediction accuracy results are plotted in Figure \ref{accuracy}A and \ref{accuracy}B. It shows that the supervised ivis framework is the best among the five dimensionality reduction methods. Surprisingly, while the unsupervised ivis model was trained without class labels in the loss function, the high prediction accuracy of this model demonstrates its good performance on the 2D projections. Random forest is often applied to distinguish the macrostates, since it provides feature importance, which is important for the interpretation of biological system. The accumulated feature importance of ranfom forest model on the supervised ivis model is plotted in Figure \ref{accuracy}C. The top $490$ features accounts for $90.2\%$ of the overall feature importance.

\begin{figure*}[t]
\centering
 \includegraphics[width=0.98\textwidth]{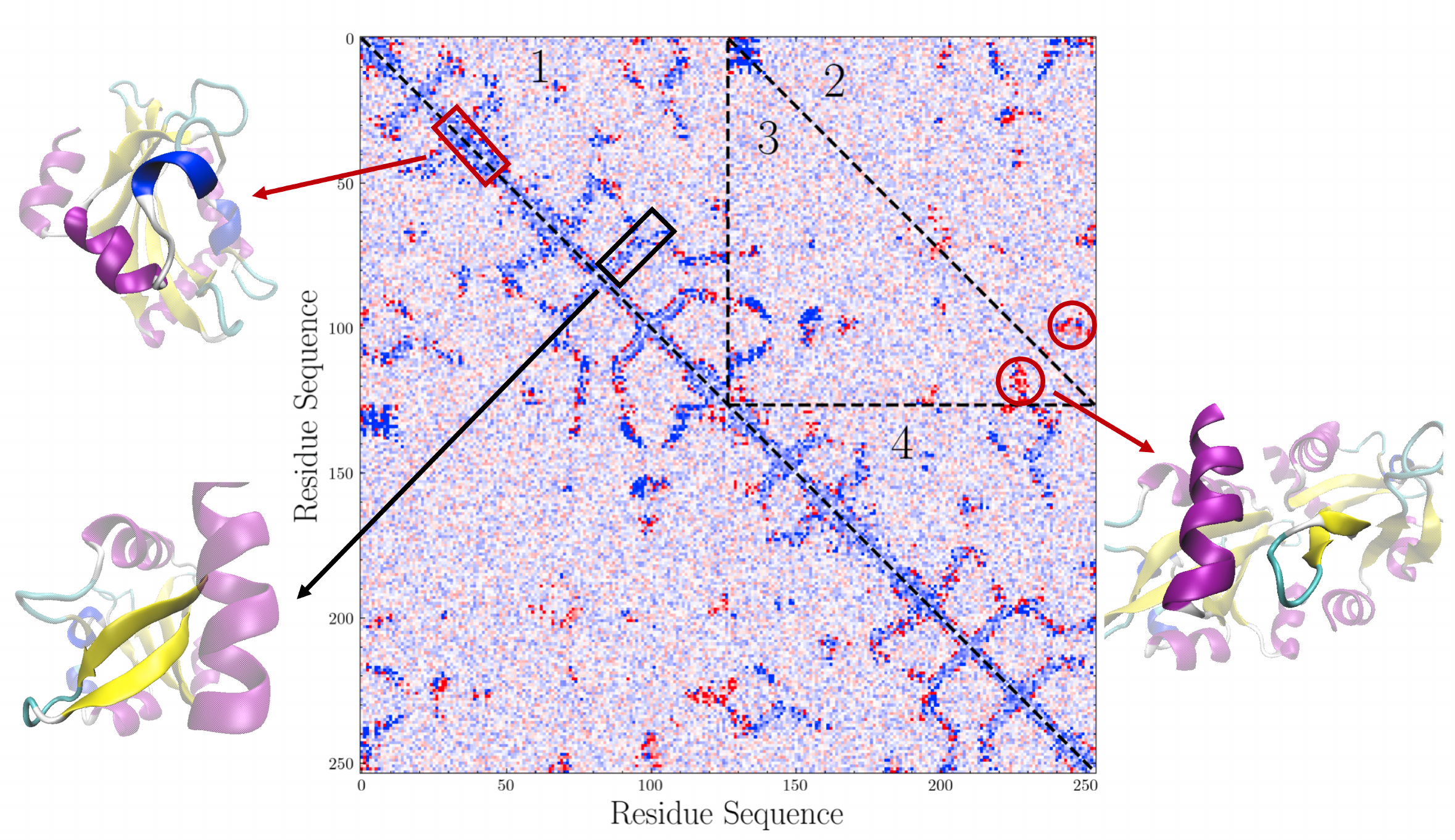}
 \caption{Protein contact map with corresponding protein structures. Feature weights of the first dense layer in the supervised ivis dimensionality reduction method were extracted and were colored as red (positive), white (close to zero) and blue (negative). The residue sequence starts from Ser240 in chain A and ends in Glu367 in chain B. }
 \label{contact}
\end{figure*}

The high prediction accuracy of the supervised ivis framework suggests that supervised ivis is more promising in elucidating the conformational differences among macrostates. The neural network architecture on the first dense layer of supervised ivis model was $32,131 \times 500$, where $32,131$ and $500$ represent the number of C$\alpha$ distances and dense layers, respectively. In order to identify key residues and structures that are important in the dimensionality reduction process, $32,131$ feature weights on the last layer were treated as the feature importance and shown as the protein contact map in Figure {\ref{contact}}. The contact map is symmetrical along the diagonal. The upper triangular part is divided into four regions as region $1$: pair wised C$\alpha$ distances within chain A, region $4$: C$\alpha$ distances within chain B, regions $2$ and $3$: C$\alpha$ distances between chain A and B. Our results show similar characteristics with a previous study {\cite{doerr2017dimensionality}}. Local protein structures are encoded to features close to the diagonal. Global structures are encoded to features further from the diagonal. In Figure {\ref{contact}}, the local information is shown in red rectangular as the C$\alpha$ and D$\alpha$ helices in AuLOV system, and global information is shown in black rectangular as the G$\beta$ and H$\beta$ strands. While region $2$ (protein interactions from chain A to chain B) and $3$ (protein interactions from chain B to chain A) are mostly symmetrical, we found the asymmetrical behavior (red circle in Figure {\ref{contact}}) that the interaction between J$\alpha$ in chain A and linkers in chain B is stronger than the interaction between J$\alpha$ in chain B and linkers in chain A.

\begin{table*}[t]
\caption{Top $20$ residues identified in the supervised ivis framework.}
\centering
\begin{tabular}{c c c c}
\hline
Residue & Importance ($\%$) & Residue & Importance ($\%$) \\
\hline
ILE 242 & $1.12$ & PHE 241& $1.10$ \\
LEU 245 & $1.07$ & \textbf{ALA 248}\textsuperscript{\emph{a}}& $1.04$ \\
\textbf{GLN 250} & $1.01$ & SER 314 & $1.00$ \\
GLN 246 & $0.99$ & \textbf{ASN 251} & $0.98$ \\
THR 247 & $0.97$ & PRO 268 & $0.97$ \\
\textbf{ASN 329} & $0.96$ & \textbf{GLN 350} & $0.96$ \\
\textbf{MET 313} & $0.95$ & ALA 244 & $0.94$ \\
\textbf{PHE 331} & $0.93$ & ASN 311 & $0.93$ \\
SER 240 & $0.92$ & \textbf{GLN 365} & $0.92$ \\
\textbf{CYS 351} & $0.91$ & ALA 335 & $0.90$ \\
\hline
\end{tabular}

\textsuperscript{\emph{a}} Experimentally confirmed important residues are shown in bold font.
\label{top 20 residues}
\end{table*}

\begin{table*}[t]
\caption{Accumulated feature importance of secondary structures. }
\centering
\begin{tabular}{c c}
\hline
Secondary structure & Importance ($\%$) \\
\hline
A'$\alpha$  &13.17\\
A$\beta$ 	&6.34\\
B$\beta$ 	&2.36\\
C$\alpha$ 	&8.50\\
D$\alpha$ 	&4.14\\
E$\alpha$ 	&1.40\\
F$\alpha$ 	&5.50\\
G$\beta$ 	&6.52\\
H$\beta$ 	&8.67\\
I$\beta$ 	&7.98\\
J$\alpha$ 	&10.44\\
Linkers 	&24.98\\
\hline
\end{tabular}
\label{structure importance}
\end{table*}

To examine the important residues identified in the protein contact map, for each C$\alpha$ distance, the corresponding feature weight was accumulated to the related two residues. Therefore, the significance of residues and structures are quantified. Top $20$ residues were listed in Table \ref{top 20 residues} with important residues that are experimentally identified {\cite{heintz2016blue, zoltowski2007conformational, moglich2009structure, herman2013blue, arinkin2017structure}} shown in bold font. The accumulated importance of secondary structure is shown in Table {\ref{structure importance}}, which shows that A'$\alpha$ helix, J$\alpha$ helix and protein linkers are important in AuLOV allostery.

\subsection*{ivis is more computationally efficient than t-ICA and t-SNE}

\begin{figure*}[t]
 \includegraphics[width=0.98\textwidth]{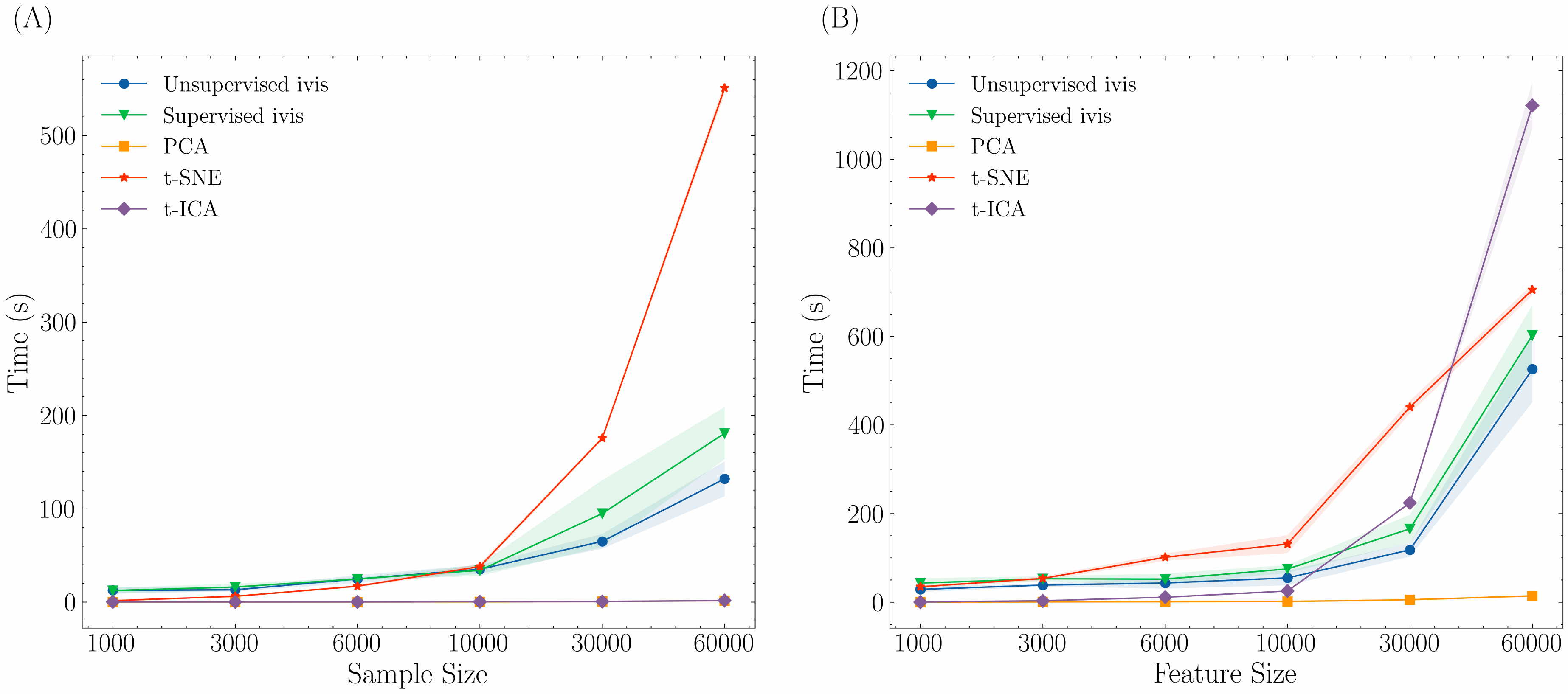}
 \caption{Computation time of each dimensionality reduction method spent for fitting high-dimensional data. (A) Runtime result of $1,000$ feature size with different sample size. Results of PCA and t-ICA were overlapped because of the timescale. (B) Runtime result of $10,000$ sample size with different feature size.}
 \label{runtime}
\end{figure*}

\begin{figure*}[t]
 \includegraphics[width=0.5\textwidth]{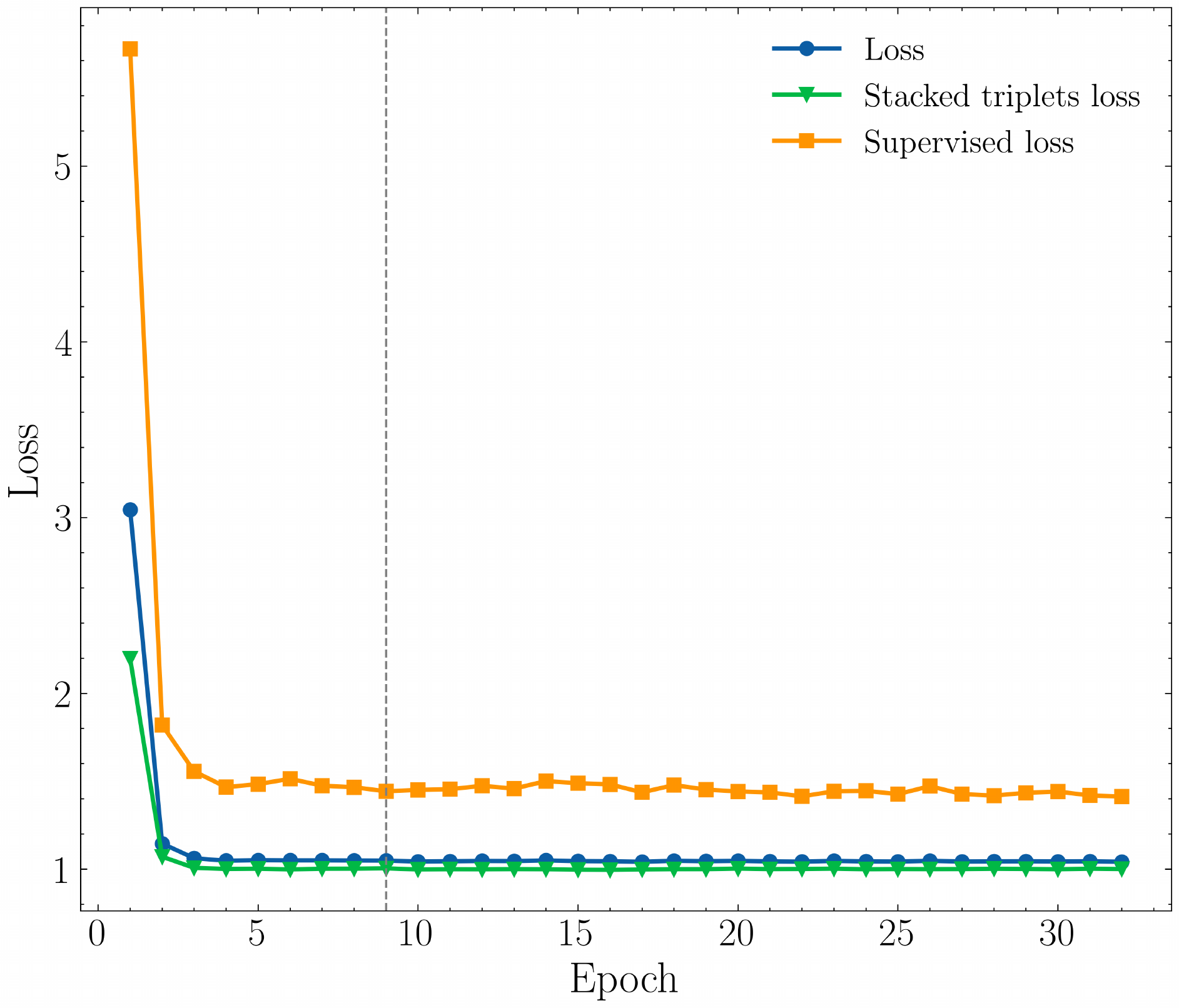}
 \caption{Triplet loss of each epoch for supervised ivis framework with supervision weight of $0.1$ and early stopping of $10$. Model is trained on dataset of $10,000$ samples with $60,000$ dimensions. Dashed grey line indicates the expected termination in training with early stopping of $5$.}
 \label{loss}
\end{figure*}

A key factor in comparing different dimensionality reduction methods is their computational cost, for it could be prohibitively expensive when dealing with large size and high-dimensional dataset. To compare the computational efficiency of different dimensionality reduction methods with regard to sample size and feature size, three randomly generated datasets with uniform distribution between $0$ and $1$ were applied for each dataset size. The relation between runtime and sample size, with feature size of $1,000$, is shown in Figure \ref{runtime}A. While t-SNE is stable and fast in small datasets ($\le 10,000$ sample size), the runtime grows the fastest among the five models and is not feasible for large dataset. t-ICA and PCA overlapped with each other since these two models are less affected by the sample size. Unsupervised ivis and supervised ivis exhibited similar runtime results. The relation between runtime and feature size with sample size of $10,000$ is shown in Figure \ref{runtime}B. t-ICA and t-SNE show similar trends in the runtime growth trend, as they perform fast in small feature size ($\le 10,000$) but not practical in higher dimensions. While both ivis models are slower than PCA, the runtime of these two models are acceptable for large sample size and high dimension. The training process of supervised ivis is further displayed in Figure \ref{loss}. Triplet loss was stable after $4$ epochs and stopped at $32$ epochs with early stopping of $10$.

\section{Discussion}

As a deep learning-based algorithm, ivis framework is originally designed for single-cell experiments to provide new approach for visualization and explanation purposes. In this study, ivis is applied on the MD simulations of allosteric protein AuLOV for dimensionality reduction. Combined with several performance criteria, ivis is demonstrated to be effective in keeping both local and global features while offering key insights for the mechanism of protein allostery. 

Various dimensionality reduction methods have been used in protein systems, such as PCA, t-ICA and t-SNE. As linear methods, PCA and t-ICA aim to capture the maximum variance and autocorrelation of protein motion, respectively. However, nonlinear dimensionality reduction methods, such as t-SNE, have been shown to be superior than linear methods in keeping the similarity between high dimension and low dimension \cite{zhou2018t}. Nevertheless, limitations of t-SNE, such as being susceptible to system noise \cite{amid2018more} and poor performance in extracting global structure, hinder further interpretations for biological systems. Compared with these dimensionality reduction methods, ivis is outstanding in preserving distances in the low-dimensional spaces and could be utilized for biological system explanation. 

Dimensionality reduction methods have different strengths in preserving structural information and can be applied to various datasets. While there is no universal standard measuring the performance of different methods, an appropriate method should reflect the distance and similarity between projections in low dimensional space. Similar structures in the high dimensional space should be close in the low dimensional space. This criterion is important in the construction of Markov state model, which requires clustering discrete microstates on the projections. Improper projections would lead to poor MSMs, thus obscuring the protein motions and hindering further structural-function study \cite{prinz2011markov}. Moreover, an adequate MSM requires the similarity between structures in each microstate. To evaluate the effectiveness of MSM, average RMSD value is often used as a good indicator for dimensionality reduction methods. With this regard, both unsupervised ivis and supervised ivis are suitable to build MSM in low dimensional space. Estimated relaxation timescale reflects the number of steady states and is used to construct kinetically stable macrostates. The timescale of protein motion ranges from milliseconds to seconds in experiments. Among all tested dimensionality reduction methods, ivis framework showed the longest timescale with over $10^{-5}$ second. This experimentally meaningful timescale, combined with the average RMSDs, suggests the success of ivis on the construction of MSM.

It is expected that Euclidean distances between data points in the high dimensional space should be proportional to the distances between the projected points in the low dimensional space. In the current study, long distance in the original dimensional space represents a high degree of dissimilarity in protein structure and the related two data points are more likely to be in different protein folding states. A well-behaved dimensionality reduction method should keep this correspondence in the low dimensional space. Several assessments are applied to quantify this relationship. Spearman's rank-order correlation coefficient is calculated to test the linear relationship of pair-wised distances of data points. A potential problem is that distances are not independent. Rather, the change in position of one residue would lead to the change in the related $n-1$ pair-wised distances. Therefore, to overcome this problem, the Mantel test is used to randomly permute rows and columns of distance matrix. The result of the Mantel test showed similar trend compared with that in Spearman correlation coefficient, which indicates that all methods are free of the dependency of distances and maintain good stability. The concept of the Shannon information in information theory is utilized to compare the information content in each projection space. The results of the above criteria show that ivis is capable of effectively separating different classes in the low dimensional space and preserve high dimensional distances with the least information loss. While high dimensional dataset is usually projected onto 2D surface, the effectiveness of MSMs on different dimensions were tested. Through the results of GMRQ, different methods showed various results. It is proposed that suitable dimensions are dependent on biological system and dimensionality reduction method. However, 2-dimension space is still desired for visualization purpose if it could represent sufficient biological information.

In the process of AuLOV dimerization, several residues have been experimentally confirmed as important in promoting the allostery. However, substantial study is necessary to establish a detailed mechanism. In the 2D projections of ivis framework, two important macrostates and corresponding protein structures can be extracted for residue-level mechanism study. The comparison of native structures reveals that it is about $7$ times more likely for the orientational changes (hydrogen bonds breaking) in Gln350 and Asn329 near cofactor FMN than the conformational changes in Phe331 and Gln350. However, because of the overlapping in the dark states, these two macrostates are missing in the projections from other dimensionality reduction methods. The protein contact map further demonstrates the superiority of ivis dimensionality reduction method that ivis can both retain local and global information. Unexpectedly, asymmetrical nature of the AuLOV dimer is revealed by comparing the protein-protein interactions. There are several important residues identified by ivis framework. Met313, Leu331 and Cys351 have been reported as light-induced rotamers near cofactor FMN \cite{heintz2016blue}. These key residues are located on the surface of the $\beta$-sheet, which is consistent with and proves the concept of signaling mechanism that signals originated from the core of Per-ARNT-Sim (PAS) generate conformational change mainly within the $\beta$-sheet \cite{zoltowski2007conformational,moglich2009structure}. Gln365 is important for the stability of J$\alpha$ helix through the hydrogen bonding with Cys316 \cite{herman2013blue}. Leu248, Gln250 and Asn251 were also found important in modulating allostery within single chain, reported as A'$\alpha$ linker while Asn329 and Gln350 function as FMN stabilizer \cite{arinkin2017structure}. Through the AuLOV dimerization, A'$\alpha$ and J$\alpha$ helices undergo conformational changes and are expected to account for large importance, as shown in Table {\ref{structure importance}}. However, the protein linkers, as well as C$\alpha$ helix and H$\beta$ and I$\beta$ strands, also showed high importance. The significance of protein linkers in the current study is consistent with both experimental and computational findings {\cite{ma2011dynamic,reddy2013linkers,george2002analysis,gokhale2000role}} that protein linkers are indispensable components in allostery and biological functions. Together, these unexpected structures are vital in AuLOV allostery and worth further study. Overall, several key residues and secondary structures identified by ivis framework agrees with the experimental finding, which consolidates the good performance of ivis in elucidating the protein allosteric process. 

Computational cost should be considered when comparing dimensionality reduction methods, since it is computationally expensive for large datasets, especially for proteins. From this prospective, different models are benchmarked using a dummy dataset. Results showed that PCA requires the least computational resource, not subjected to either sample size or feature size. This might be due to the reason that PCA implemented in Scikit-learn uses SVD for acceleration. Further, since the size of dataset was large, randomized truncated SVD was applied to reduce the time complexity to $\mathcal{O}(n_{\text{max}}^2 \cdot n_{\text{components}})$ with $n_{\text{max}} = \max (n_{\text{samples}}, n_{\text{features}})$ \cite{halko2009finding}. While t-SNE is comparable with ivis regarding several assessments, the computational cost could be prohibitively expensive for large datasets as t-SNE has a time complexity of $\mathcal{O}(N^2D)$ \cite{ding2018interpretable}, where $N$ and $D$ are the number of samples and features, respectively. Though tree-based algorithms have been developed to reduce the complexity to $\mathcal{O}(N \log N)$ \cite{van2014accelerating}, it is still challenging for the high-dimensional protein system. ivis exhibited less computational cost in higher sample size and dimension. Further, as shown in Figure \ref{loss}, the loss of ivis model converges fast and the overall computational cost could have been further reduced with early stopping iterations. Combined with the performance criteria and runtime comparison, ivis framework is demonstrated as a superior dimensionality reduction method for protein system and can be an important member in the analysis toolbox for MD trajectory.

\section{Conclusion}

As originally developed for single-cell technology, ivis framework is applied in this study as a dimensionality reduction method for molecular dynamics simulations for biological macromolecules. ivis is superior than other dimensionality reduction methods in several aspects, ranging from preserving both local and global distances, maintaining similarity among data points in high dimensional space and projections, to retaining the most structural information through a series of performance assessments. ivis also shows great potential in interpreting biological system through the feature weights in the neural network layer. Overall, ivis reached a balance between dimensionality reduction performance and computational cost and is therefore promising as an effective tool for the analysis of macromolecular simulations.

\begin{acknowledgement}
Research reported in this paper was supported by the National Institute of General Medical Sciences of the National Institutes of Health under Award No. R15GM122013. Computational time was generously provided by Southern Methodist University's Center for Research Computing. The authors thank Miss Xi Jiang from Biostatistics Ph.D. program in the Statistics department of SMU for fruitful discussions. 
\end{acknowledgement}

%\onecolumn
%\newpage
\bibliography{ref-abbr, ref}

%\begin{suppinfo}
%See supporting information for additional experimental details.
%\end{suppinfo}

\end{document}